\documentclass[11pt]{article}
\usepackage{amsmath}
\usepackage{amsfonts}
\usepackage[a4paper]{geometry}

\input{tcilatex}
\begin{document}

\title{Repo convexity}
\author{Paul McCloud \\
Department of Mathematics, University College London}
\maketitle

\let\thefootnote\relax\footnote{Author email: p.mccloud@ucl.ac.uk}

\begin{abstract}
There is an observed basis between repo discounting, implied from market
repo rates, and bond discounting, stripped from the market prices of the
underlying bonds. Here, this basis is explained as a convexity effect
arising from the decorrelation between the discount rates for derivatives
and bonds.

Using a Hull-White model for the discount basis, expressions are derived
that can be used to interpolate the repo rates of bonds with different
maturities and to extrapolate the repo curve for discounting
bond-collateralised derivatives.\bigskip


\end{abstract}

\newpage

\section{Securities and derivatives}

No-arbitrage assumptions imply the fair value of a payoff is determined from
a risk-neutral pricing measure $\mathbb{E}$ and a predictable numeraire $%
q_{t}$ via the martingale condition:%
\begin{equation}
\frac{a_{t}}{q_{t}}=\mathbb{E}_{t}[\frac{a_{T}}{q_{T}}]
\end{equation}%
While this model appears to be sensitive to the numeraire, in practice the
only property required of the numeraire is predictability. Securities with
observable market prices are marked-to-market, while the inclusion of
funding flows implies the discounting of derivatives follows the discounting
of the securities used to fund them. This argument is considered in more
detail below.

Consider a security with observable market price $\bar{p}_{t}$. The discount
rate $\bar{r}_{t}$ associated with the security is defined to be the
risk-neutral expected return on the security:%
\begin{equation}
\bar{r}_{t}\,dt=\frac{\mathbb{E}_{t}[d\bar{p}_{t}]}{\bar{p}_{t}}
\end{equation}%
Dividends, which are excluded in this model for the security, can be added
as discrete terms in the expected return. The expression for the discount
rate is integrated to the martingale property:%
\begin{equation}
\bar{p}_{t}=\mathbb{E}_{t}[\exp [-\int_{\tau =t}^{T}\bar{r}_{\tau }\,d\tau ]%
\bar{p}_{T}]
\end{equation}%
allowing the security price to be modelled in terms of the terminal payoff
and the security discount rate $\bar{r}_{t}$.

Consider a derivative with price $a_{t}$ that can be funded with any of a
range of securities with prices $\bar{p}_{t}^{i}$. Funding with the $i$th
security, the return to the seller of the derivative over the interval $dt$
is $(a_{t}/\bar{p}_{t}^{i})d\bar{p}_{t}^{i}-da_{t}$. The seller chooses to
invest the proceeds of the sale in the funding security that maximises the
return value. The fair value model then implies the price expression:%
\begin{align}
0& =\max_{i}\mathbb{E}_{t}[\frac{(a_{t}/\bar{p}_{t}^{i})d\bar{p}%
_{t}^{i}-da_{t}}{q_{t}+dq_{t}}] \\
& =\frac{a_{t}\max_{i}[\bar{r}_{t}^{i}]\,dt-\mathbb{E}_{t}[da_{t}]}{%
q_{t}+dq_{t}}  \notag
\end{align}%
where predictability has been used to take the numeraire outside the
expectation. The numeraire can now be cancelled, and the expression
integrated to generate the martingale property:%
\begin{equation}
a_{t}=\mathbb{E}_{t}[\exp [-\int_{\tau =t}^{T}r_{\tau }\,d\tau ]a_{T}]
\end{equation}%
allowing the derivative price to be modelled in terms of the terminal payoff
and the derivative discount rate $r_{t}$:%
\begin{equation}
r_{t}=\max_{i}[\bar{r}_{t}^{i}]
\end{equation}%
This argument assumes that the funding security can be switched in its
entirety at any time, which is not typically the case. More generally, the
relationship between the derivative discount rate and the discount rates of
the securities that fund it is heavily path-dependent.

As a special case, the derivative discount factor for maturity $T$ is:%
\begin{equation}
p_{t}^{T}=\mathbb{E}_{t}[\exp [-\int_{\tau =t}^{T}r_{\tau }\,d\tau ]]
\end{equation}%
being the price of the derivative with unit payoff at maturity.

The martingale expressions for the security and derivative prices involve
different discount rates but are otherwise the same. In the following, the
volatility of the difference between these discount rates is used to explain
the observed basis between the repo discount factors implied from observed
repo rates and the discount factors stripped from the bond curve.

\section{Repo convexity}

The bond market for an issuer is assumed to comprise discount bonds with
price $\bar{p}_{t}^{T}$ for each maturity $T$. Ignoring the possibility of
default, the bond satisfies the boundary condition $\bar{p}_{T}^{T}=1$, and
the martingale property for the bond price becomes:%
\begin{equation}
\bar{p}_{t}^{T}=\mathbb{E}_{t}[\exp [-\int_{\tau =t}^{T}\bar{r}_{\tau
}^{T}\,d\tau ]]
\end{equation}%
Common features among the family of bonds that derive from macro-economic
considerations and the conditions of the issuer are encapsulated in the
curve contribution $\bar{r}_{t}$ to the bond discount rate, with the
residual contribution $\bar{z}_{t}^{T}$ for the individual bond reflecting
liquidity and investor preference. The bond discount rate $\bar{r}_{t}^{T}$
then decomposes as:%
\begin{equation}
\bar{r}_{t}^{T}=\bar{r}_{t}+\bar{z}_{t}^{T}
\end{equation}%
There is a degree of arbitrariness in this decomposition, and expert
knowledge is required to separate curve and liquidity contributions to the
discount rate.

Consider a forward-starting repo that sets at time $t$ over the period
starting at time $s$ and ending at time $e$ on the bond maturing at time $T$%
, where $t\leq s<e\leq T$. At time $s$, the unit cashflow is exchanged for $%
(1/\bar{p}_{s}^{T})$ units of the bond, a price-neutral exchange. At time $e$%
, the bonds are returned in exchange for the cashflow $(1+f_{ts}^{eT}\delta
) $, where $f_{ts}^{eT}$ is the repo rate and $\delta $ is the daycount.
Haircuts and bond coupons are not considered in this construction, though
both features are straightforward to add albeit at the cost of additional
complexity in the expression for the repo rate.

By construction this has zero price at time $t$, leading to the price
expression:%
\begin{equation}
0=\mathbb{E}_{t}[\exp [-\int_{\tau =t}^{e}r_{\tau }\,d\tau ](\frac{\bar{p}%
_{e}^{T}}{\bar{p}_{s}^{T}}-(1+f_{ts}^{eT}\delta ))]
\end{equation}%
Repo convexity arises from the dual-discounted nature of this construction.
The repo price depends on the bond discount rate $\bar{r}_{t}^{T}$ that
determines the bond price at settlement and the derivative discount rate $%
r_{t}$ used for discounting. Decorrelation between these discount rates
leads to a convexity adjustment for the repo rate.

The repo price expression can be re-arranged to identify the convexity
adjustment for the repo rate. Define the discounting basis $b_{t}^{T}$ and
the liquidity basis $\bar{s}_{t}^{eT}$:%
\begin{align}
b_{t}^{T}& =r_{t}-\bar{r}_{t}^{T} \\
\bar{s}_{t}^{eT}& =\bar{z}_{t}^{T}-\bar{z}_{t}^{e}  \notag
\end{align}%
First note that:%
\begin{align}
\mathbb{E}_{t}[& \exp [-\int_{\tau =t}^{e}r_{\tau }\,d\tau ]\frac{\bar{p}%
_{e}^{T}}{\bar{p}_{s}^{T}}] \\
& =\mathbb{E}_{t}[\exp [-\int_{\tau =t}^{e}\bar{r}_{\tau }^{T}\,d\tau ]\exp
[-\int_{\tau =t}^{e}b_{\tau }^{T}\,d\tau ]\frac{\bar{p}_{e}^{T}}{\bar{p}%
_{s}^{T}}]  \notag \\
& =\dfrac{\mathbb{E}_{t}[\exp [-\int_{\tau =t}^{e}\bar{r}_{\tau }^{T}\,d\tau
]\exp [-\int_{\tau =t}^{e}b_{\tau }^{T}\,d\tau ]]\mathbb{E}_{t}[\exp
[-\int_{\tau =t}^{e}\bar{r}_{\tau }^{T}\,d\tau ](\bar{p}_{e}^{T}/\bar{p}%
_{s}^{T})]}{\mathbb{E}_{t}[\exp [-\int_{\tau =t}^{e}\bar{r}_{\tau
}^{T}\,d\tau ]]}\exp [C_{ts}^{eT}]  \notag \\
& =\mathbb{E}_{t}[\exp [-\int_{\tau =t}^{e}r_{\tau }\,d\tau ]]\dfrac{\mathbb{%
E}_{t}[\exp [-\int_{\tau =t}^{s}\bar{r}_{\tau }^{T}\,d\tau ]]}{\mathbb{E}%
_{t}[\exp [-\int_{\tau =t}^{e}\bar{r}_{\tau }^{T}\,d\tau ]]}\exp
[C_{ts}^{eT}]  \notag \\
& =p_{t}^{e}\frac{\bar{p}_{t}^{s}\exp [-L_{t}^{sT}]}{\bar{p}_{t}^{e}\exp
[-L_{t}^{eT}]}\exp [C_{ts}^{eT}]  \notag
\end{align}%
The liquidity adjustment $L_{t}^{eT}$ and convexity adjustment $C_{ts}^{eT}$
in this expression are:%
\begin{align}
L_{t}^{eT}& =\log [\frac{\mathbb{E}_{t}[\exp [-\int_{\tau =t}^{e}\bar{r}%
_{\tau }^{e}\,d\tau ]]}{\mathbb{E}_{t}[\exp [-\int_{\tau =t}^{e}\bar{r}%
_{\tau }^{T}\,d\tau ]]}] \\
C_{ts}^{eT}& =\log [\frac{\mathbb{E}_{t}[\exp [-\int_{\tau =t}^{e}\bar{r}%
_{\tau }^{T}\,d\tau ]]\mathbb{E}_{t}[\exp [-\int_{\tau =t}^{e}\bar{r}_{\tau
}^{T}\,d\tau ]\exp [-\int_{\tau =t}^{e}b_{\tau }^{T}\,d\tau ](\bar{p}%
_{e}^{T}/\bar{p}_{s}^{T})]}{\mathbb{E}_{t}[\exp [-\int_{\tau =t}^{e}\bar{r}%
_{\tau }^{T}\,d\tau ]\exp [-\int_{\tau =t}^{e}b_{\tau }^{T}\,d\tau ]]\mathbb{%
E}_{t}[\exp [-\int_{\tau =t}^{e}b_{\tau }^{T}\,d\tau ](\bar{p}_{e}^{T}/\bar{p%
}_{s}^{T})]}]  \notag
\end{align}%
The recurrence of the integral kernel:%
\begin{equation}
\exp [-\int_{\tau =t}^{e}\bar{r}_{\tau }^{T}\,d\tau ]
\end{equation}%
in these expressions suggests the switch to the equivalent measure $\mathbb{%
\bar{E}}^{eT}$ related to the risk-neutral measure $\mathbb{E}$ by the
Radon-Nikodym derivative:%
\begin{equation}
\frac{d\mathbb{\bar{E}}^{eT}}{d\mathbb{E}}=\frac{\exp [-\int_{\tau =0}^{e}%
\bar{r}_{\tau }^{T}\,d\tau ]}{\mathbb{E}[\exp [-\int_{\tau =0}^{e}\bar{r}%
_{\tau }^{T}\,d\tau ]]}
\end{equation}%
In this measure, the liquidity and convexity adjustments simplify:%
\begin{align}
L_{t}^{eT}& =\log [\mathbb{\bar{E}}_{t}^{eT}[\exp [\int_{\tau =t}^{e}\bar{s}%
_{\tau }^{eT}\,d\tau ]]] \\
C_{ts}^{eT}& =\log [\frac{\mathbb{\bar{E}}_{t}^{eT}[\exp [-\int_{\tau
=t}^{e}b_{\tau }^{T}\,d\tau ](\bar{p}_{e}^{T}/\bar{p}_{s}^{T})]}{\mathbb{%
\bar{E}}_{t}^{eT}[\exp [-\int_{\tau =t}^{e}b_{\tau }^{T}\,d\tau ]]\mathbb{%
\bar{E}}_{t}^{e}[\bar{p}_{e}^{T}/\bar{p}_{s}^{T}]}]  \notag
\end{align}%
demonstrating that $L_{t}^{eT}$ is driven by the liquidity basis and $%
C_{ts}^{eT}$ is driven by the covariance between the bond discount rate and
the discount basis.

Further decompose the convexity adjustment into the maturity adjustment $%
M_{t}^{eT}$ and the forwardness adjustment $F_{ts}^{eT}$:%
\begin{align}
M_{t}^{eT}& =C_{tt}^{eT} \\
F_{ts}^{eT}& =C_{ts}^{eT}-(C_{tt}^{eT}-C_{tt}^{sT})  \notag
\end{align}%
so that:%
\begin{equation}
C_{ts}^{eT}=(M_{t}^{eT}-M_{t}^{sT})+F_{ts}^{eT}
\end{equation}%
The expression for the repo rate is then:%
\begin{equation}
f_{ts}^{eT}=\frac{1}{\delta }(\frac{\hat{p}_{t}^{sT}}{\hat{p}_{t}^{eT}}\exp
[F_{ts}^{eT}]-1)
\end{equation}%
where the repo discount factor $\hat{p}_{t}^{eT}$ is defined by:%
\begin{equation}
\hat{p}_{t}^{eT}=\bar{p}_{t}^{e}\exp [-L_{t}^{eT}-M_{t}^{eT}]
\end{equation}%
This shows that the repo rate follows the standard formula in terms of the
bond discount factors, with liquidity and maturity adjustments applied to
the discount factors and forwardness adjustment applied to the rate.

The liquidity adjustment $L_{t}^{eT}$ is the adjustment applied to the repo
discount factor $\hat{p}_{t}^{eT}$ to account for the liquidity spread
between the bond maturing at time $e$ and the bond maturing at time $T$. The
liquidity adjustment depends on the mean and variance of the liquidity basis:%
\begin{equation}
L_{t}^{eT}=\log [\mathbb{\bar{E}}[\exp [S]]]\approx \mu _{S}+\frac{1}{2}%
\sigma _{S}^{2}
\end{equation}%
where:%
\begin{equation}
S=\int_{\tau =t}^{e}\bar{s}_{\tau }^{eT}\,d\tau
\end{equation}%
The approximation is exact when the variable $S$ is normal in the measure $%
\mathbb{\bar{E}}\equiv \mathbb{\bar{E}}_{t}^{eT}$.

The maturity adjustment $M_{t}^{eT}$ is the convexity adjustment applied to
the repo discount factor $\hat{p}_{t}^{eT}$ to account for the delay between
the settlement of the repo and the maturity of the bond. This adjustment
satisfies the boundary condition $M_{t}^{ee}=0$, so that there is no
maturity adjustment in the repo-to-maturity case $e=T$. The forwardness
adjustment $F_{ts}^{eT}$ is the convexity adjustment applied to the repo
rate $f_{ts}^{eT}$ to account for the delay between the fixing of the repo
rate and the start of the repo period. This adjustment satisfies the
boundary condition $F_{tt}^{eT}=0$, so there is no forwardness adjustment in
the spot-starting case $t=s$. The convexity adjustment depends on the
covariance between the discount basis and the bond price:%
\begin{equation}
C_{ts}^{eT}=\log [\frac{\mathbb{\bar{E}}[\exp [-B+P]]}{\mathbb{\bar{E}}[\exp
[-B]]\mathbb{\bar{E}}[\exp [P]]}]\approx -\rho _{BP}\sigma _{B}\sigma _{P}
\end{equation}%
where:%
\begin{align}
B& =\int_{\tau =t}^{e}b_{\tau }^{T}\,d\tau \\
P& =\log [\frac{\bar{p}_{e}^{T}}{\bar{p}_{s}^{T}}]  \notag
\end{align}%
The approximation is exact when the variables $B$ and $P$ are joint normal
in the measure $\mathbb{\bar{E}}\equiv \mathbb{\bar{E}}_{t}^{eT}$.

\section{Hull-White model for repo convexity}

In this section, the liquidity contribution to the bond discount rate is
taken to be zero, and a model for the convexity adjustment is constructed
using correlated Hull-White models for the bond discount rate and the
discount basis. The convexity adjustment depends on the correlations between
the variables:%
\begin{align}
R& =\int_{\tau =t}^{e}\bar{r}_{\tau }\,d\tau \\
B& =\int_{\tau =t}^{e}b_{\tau }\,d\tau  \notag \\
P& =\log [\dfrac{\bar{p}_{e}^{T}}{\bar{p}_{s}^{T}}]  \notag
\end{align}%
When these variables are joint normal in the risk-neutral measure $\mathbb{E}
$ the convexity adjustment becomes:%
\begin{equation}
C_{ts}^{eT}=-\rho _{BP}\sigma _{B}\sigma _{P}
\end{equation}%
where $\sigma _{B}$ and $\sigma _{P}$ are the standard deviations of $B$ and 
$P$ and $\rho _{BP}$ is the correlation between them.

In order to generate an expression for the convexity adjustment, consider
the Hull-White model for the bond discount rate and the discount basis:%
\begin{align}
d\bar{r}_{t}& =\theta (\bar{r}_{t}^{\ast }-\bar{r}_{t})\,dt+\sigma \,dx_{t}
\\
db_{t}& =\kappa (b_{t}^{\ast }-b_{t})\,dt+\varepsilon \,dy_{t}  \notag
\end{align}%
where $\sigma $ and $\varepsilon $ are the normal volatilities and $\theta $
and $\kappa $ are the mean reversion rates of the bond discount rate and
discount basis, and the Brownian processes $x_{t}$ and $y_{t}$, driftless in
the risk-neutral measure, are correlated:%
\begin{equation}
dx_{t}\,dy_{t}=\rho \,dt
\end{equation}%
The mean reversion levels $\bar{r}_{t}^{\ast }$ and $b_{t}^{\ast }$ are
calibrated to the initial bond and derivative discount factors.

The core variables whose covariance generates the convexity are normal in
this model. Integrating the model leads to:%
\begin{align}
\log [\bar{p}_{s}^{T}]& =-\frac{\sigma }{\theta }(1-e^{-\theta
(T-s)})\int_{\tau =t}^{s}e^{-\theta (s-\tau )}\,dx_{\tau }+\text{drift} \\
\log [\bar{p}_{e}^{T}]& =-\frac{\sigma }{\theta }(1-e^{-\theta
(T-e)})\int_{\tau =t}^{e}e^{-\theta (e-\tau )}\,dx_{\tau }+\text{drift} 
\notag \\
\int_{\tau =t}^{e}b_{\tau }\,d\tau & =\frac{\varepsilon }{\kappa }\int_{\tau
=t}^{e}(1-e^{-\kappa (e-\tau )})\,dy_{\tau }+\text{drift}  \notag
\end{align}%
The covariance that generates the convexity adjustment is then:%
\begin{align}
C_{ts}^{eT}=\frac{\rho \sigma \varepsilon }{\theta \kappa }& ((1-e^{-\theta
(T-e)})\int_{\tau =t}^{e}e^{-\theta (e-\tau )}(1-e^{-\kappa (e-\tau
)})\,d\tau \\
& -(1-e^{-\theta (T-s)})\int_{\tau =t}^{s}e^{-\theta (s-\tau )}(1-e^{-\kappa
(e-\tau )})\,d\tau )  \notag
\end{align}%
The convexity depends on the three time intervals:%
\begin{align}
\tau & =s-t \\
\delta & =e-s  \notag \\
\mu & =T-e  \notag
\end{align}%
where $\tau $ is the forwardness of the repo, $\delta $ is the length of the
repo period, and $\mu $ is the time-to-maturity from the end of the repo of
the reference discount bond. The convexity can then be expressed as:%
\begin{equation}
C_{ts}^{eT}=\rho \sigma \varepsilon B[s-t,e-s,T-e;\theta ,\kappa ]
\end{equation}%
where:%
\begin{align}
B[\tau ,\delta ,\mu ;\theta ,\kappa ]=\,& \frac{1}{\theta \kappa }%
(1-e^{-\theta \mu })(\frac{1}{\theta }(1-e^{-\theta (\tau +\delta )})-\frac{1%
}{\theta +\kappa }(1-e^{-(\theta +\kappa )(\tau +\delta )})) \\
& -\frac{1}{\theta \kappa }(1-e^{-\theta (\delta +\mu )})(\frac{1}{\theta }%
(1-e^{-\theta \tau })-\frac{1}{\theta +\kappa }e^{-\kappa \delta
}(1-e^{-(\theta +\kappa )\tau }))  \notag
\end{align}%
This function has finite limits as $\theta $ and $\kappa $ tend to zero. The
maturity and forwardness adjustments are then:%
\begin{align}
M_{t}^{eT}& =\frac{\rho \sigma \varepsilon }{\theta \kappa }(1-e^{-\theta
(T-e)})(\frac{1}{\theta }(1-e^{-\theta (e-t)})-\frac{1}{\theta +\kappa }%
(1-e^{-(\theta +\kappa )(e-t)})) \\
F_{ts}^{eT}& =-\frac{\rho \sigma \varepsilon }{\theta \kappa (\theta +\kappa
)}(1-e^{-\theta (T-s)})(1-e^{-\kappa (e-s)})(1-e^{-(\theta +\kappa )(s-t)}) 
\notag
\end{align}

\section{Calibration to repo discount factors}

The convexity adjustment satisfies the boundary condition $C_{ss}^{ee}=0$ in
the spot-starting repo-to-maturity case, in which case the repo rate is:%
\begin{equation}
f_{ss}^{ee}=\frac{1}{\delta }(\frac{1}{\bar{p}_{s}^{e}}-1)
\end{equation}%
More generally, the repo rate is impacted by the convexity adjustment $%
C_{ts}^{eT}$ due to the time-to-maturity $\mu =T-e$ of the reference bond
and the forwardness $\tau =s-t$ of the repo.

Consider first the convexity adjustment arising from the bond maturity. In
the spot-starting infinite maturity case the repo rate is given by:%
\begin{equation}
f_{ss}^{e\infty }=\frac{1}{\delta }(\frac{1}{\hat{p}_{s}^{e}}-1)
\end{equation}%
where the repo discount factor $\hat{p}_{t}^{e}$ is defined by:%
\begin{equation}
\hat{p}_{t}^{e}=\bar{p}_{t}^{e}\exp [-\frac{\rho \sigma \varepsilon }{\theta
\kappa }(\frac{1}{\theta }(1-e^{-\theta (e-t)})-\frac{1}{\theta +\kappa }%
(1-e^{-(\theta +\kappa )(e-t)}))]
\end{equation}%
This expression defines the repo discount factor in terms of the bond
discount factor and the model parameters. The instantaneous forward rates $%
\bar{f}_{t}^{e}$ and $\hat{f}_{t}^{e}$ for the bond and repo are then
related by:%
\begin{equation}
\hat{f}_{t}^{e}=\bar{f}_{t}^{e}+\frac{\rho \sigma \varepsilon }{\theta
\kappa }e^{-\theta (e-t)}(1-e^{-\kappa (e-t)})
\end{equation}%
If the repo forward rates are observed in the market up to some finite
maturity $E$, this model can be used to extrapolate the repo curve using the
bond forward rates as reference:%
\begin{equation}
\hat{f}_{t}^{e}=\bar{f}_{t}^{e}+(\hat{f}_{t}^{E}-\bar{f}_{t}^{E})e^{-\theta
(e-E)}\frac{1-e^{-\kappa (e-t)}}{1-e^{-\kappa (E-t)}}
\end{equation}%
Practical applications of this expression include the extrapolation of the
repo curve for use in discounting bond-collateralised derivatives.

Extending to the spot-starting finite maturity case, the repo rate is:%
\begin{equation}
f_{ss}^{eT}=\frac{1}{\delta }(\frac{1}{\hat{p}_{s}^{eT}}-1)
\end{equation}%
where the repo discount factor $\hat{p}_{t}^{eT}$ geometrically interpolates
between the bond discount factor $\bar{p}_{t}^{e}$ and the repo discount
factor $\hat{p}_{t}^{e}$:%
\begin{equation}
\hat{p}_{t}^{eT}=(\bar{p}_{t}^{e})^{\exp [-\theta (T-e)]}(\hat{p}%
_{t}^{e})^{1-\exp [-\theta (T-e)]}
\end{equation}%
The convexity adjustment for the spot-starting repo is absorbed in the
definition of the repo discount factors. The only model parameter that
appears in this expression is the mean reversion rate for the bond discount
rate, which determines the speed of interpolation between the bond and repo
discount factors as the maturity increases.

The convexity adjustment arising from the forwardness of the repo cannot be
absorbed as an adjustment to the repo discount factors. Including the
forwardness adjustment, the general expression for the repo rate is:%
\begin{equation}
f_{ts}^{eT}=\frac{1}{\delta }(\frac{\hat{p}_{t}^{sT}}{\hat{p}_{t}^{eT}}\exp
[-\frac{\rho \sigma \varepsilon }{\theta \kappa (\theta +\kappa )}%
(1-e^{-\theta (T-s)})(1-e^{-\kappa (e-s)})(1-e^{-(\theta +\kappa )(s-t)})]-1)
\end{equation}%
This model geometrically interpolates between the zero and infinite
forwardness cases:%
\begin{equation}
1+f_{ts}^{eT}\delta =(1+f_{ss}^{eT}\delta )^{\exp [-(\theta +\kappa
)(s-t)]}(1+f_{-\infty s}^{eT}\delta )^{1-\exp [-(\theta +\kappa )(s-t)]}
\end{equation}%
The contribution from forwardness is implemented as a convexity adjustment
to the ratio of repo discount factors. This convexity adjustment is nonzero
even in the repo-to-maturity case when the repo is forward-starting.

\end{document}